\documentclass[fleqn,usenatbib,letterpaper]{mn2e}
\usepackage{graphicx}
\usepackage{amsmath}
\usepackage{array}
\newcolumntype{L}[1]{>{\raggedright\let\newline\\\arraybackslash\hspace{0pt}}m{#1}}

\def\jnl@style{\it}
\def\aaref@jnl#1{{\jnl@style#1}}

\def\aaref@jnl#1{{\jnl@style#1}}

\def\aj{\aaref@jnl{AJ}}                   
\def\araa{\aaref@jnl{ARA\&A}}             
\def\apj{\aaref@jnl{ApJ}}                 
\def\apjl{\aaref@jnl{ApJ}}                
\def\apjs{\aaref@jnl{ApJS}}               
\def\ao{\aaref@jnl{Appl.~Opt.}}           
\def\apss{\aaref@jnl{Ap\&SS}}             
\def\aap{\aaref@jnl{A\&A}}                
\def\aapr{\aaref@jnl{A\&A~Rev.}}          
\def\aaps{\aaref@jnl{A\&AS}}              
\def\azh{\aaref@jnl{AZh}}                 
\def\baas{\aaref@jnl{BAAS}}               
\def\jrasc{\aaref@jnl{JRASC}}             
\def\memras{\aaref@jnl{MmRAS}}            
\def\mnras{\aaref@jnl{MNRAS}}             
\def\pra{\aaref@jnl{Phys.~Rev.~A}}        
\def\prb{\aaref@jnl{Phys.~Rev.~B}}        
\def\prc{\aaref@jnl{Phys.~Rev.~C}}        
\def\prd{\aaref@jnl{Phys.~Rev.~D}}        
\def\pre{\aaref@jnl{Phys.~Rev.~E}}        
\def\prl{\aaref@jnl{Phys.~Rev.~Lett.}}    
\def\pasp{\aaref@jnl{PASP}}               
\def\pasj{\aaref@jnl{PASJ}}               
\def\qjras{\aaref@jnl{QJRAS}}             
\def\skytel{\aaref@jnl{S\&T}}             
\def\solphys{\aaref@jnl{Sol.~Phys.}}      
\def\sovast{\aaref@jnl{Soviet~Ast.}}      
\def\ssr{\aaref@jnl{Space~Sci.~Rev.}}     
\def\zap{\aaref@jnl{ZAp}}                 
\def\nat{\aaref@jnl{Nature}}              
\def\iaucirc{\aaref@jnl{IAU~Circ.}}       
\def\aplett{\aaref@jnl{Astrophys.~Lett.}} 
\def\apspr{\aaref@jnl{Astrophys.~Space~Phys.~Res.}}
\def\bain{\aaref@jnl{Bull.~Astron.~Inst.~Netherlands}} 
\def\fcp{\aaref@jnl{Fund.~Cosmic~Phys.}}  
\def\gca{\aaref@jnl{Geochim.~Cosmochim.~Acta}}   
\def\grl{\aaref@jnl{Geophys.~Res.~Lett.}} 
\def\jcp{\aaref@jnl{J.~Chem.~Phys.}}      
\def\jgr{\aaref@jnl{J.~Geophys.~Res.}}    
\def\jqsrt{\aaref@jnl{J.~Quant.~Spec.~Radiat.~Transf.}}
\def\memsai{\aaref@jnl{Mem.~Soc.~Astron.~Italiana}}
\def\nphysa{\aaref@jnl{Nucl.~Phys.~A}}   
\def\physrep{\aaref@jnl{Phys.~Rep.}}   
\def\physscr{\aaref@jnl{Phys.~Scr}}   
\def\planss{\aaref@jnl{Planet.~Space~Sci.}}   
\def\procspie{\aaref@jnl{Proc.~SPIE}}   

\begin{document}

\title[Plan $\beta$: Core or Cusp?]
{Plan $\beta$: Core or Cusp?}

\author[T. D. Richardson, D.Spolyar \& \ M. D. Lehnert]{Thomas Richardson\thanks{thomas.d.richardson@kcl.ac.uk}$^{1}$, Douglas Spolyar\thanks{dspolyar@gmail.com}$^{2,3}$ \&
Matthew D. Lehnert\thanks{lehnert@iap.fr}$^{2}$
\\$^{1}$Physics, Kings College London, Strand, London WC2R 2LS, UK\\$^{2}$Institut d'Astrophysique de Paris - 98 bis boulevard Arago - 75014 Paris, France
\\$^3$GRAPPA Institute, University of Amsterdam, Science Park 904, 1090 GL Amsterdam, The Netherlands}
\maketitle
\begin{abstract}

The inner profile of Dark Matter (DM) halos remains one of the central problems in small-scale cosmology. At present, the problem can not be resolved in dwarf spheroidal galaxies due to a degeneracy between the DM profile and the velocity anisotropy $\beta$ of the stellar population. We discuss a method which can break the degeneracy by exploiting 3D positions and 1D line-of-sight (LOS) velocities.
With the full 3D spatial information, we can determine precisely what fraction of each stars LOS motion is in the radial and tangential direction. This enables us to infer the anisotropy parameter $\beta$ directly from the data. The method is particularly effective if the galaxy is highly anisotropic.  
Finally, we argue that such a test could be applied to Sagittarius and potentially other dwarfs with RR Lyrae providing the necessary depth information.
\end{abstract}


\section{Introduction}

A central prediction of Cold Dark Matter (CDM) simulations \citep{NFWhalos} is the formation of a cusp at the center of a DM halo. Mounting (though still debated) observational evidence (See  \cite{spiralcores} for spiral galaxies) contrarily prefers a cored halo.  This issue has been dubbed the ``cusp vs core problem".
 If indeed halos are cored, then our present understanding of CDM halos requires modification from either astrophysical effects or new particle physics. On the astrophysical side, numerous explanations \citep{astrocore,nomorecusp,astrocorefeedback} have been given that may reduce the cusp of a halo. 
On the other hand, if DM has an internal force, then the self interaction \citep{firstSI} can induce a core \citep{walkSI}. 
 
At a practical level, the DM profile is also an important uncertainty for indirect detection experiments.
   Dwarf galaxies are a primary target for indirect experiments e.g the FERMI gamma ray telescope \citep{fermidwarf}  
   looking for a DM annihilation signal.  The potential signal  is highly sensitive to the DM profile with
   the annihilation rate going like the density squared $\sim \rho^2$. 
Reducing this astrophysical uncertainty will improve indirect detection constraints on the particle physics properties of DM.
     Hence, determining the density profile of a dwarf galaxy
is a critical bench mark in deciphering the nature of dark matter and structure formation in general.  

Unfortunately the density of dark matter at the center of distant spherical systems is very poorly constrained by the simplest and most widely used method in galactic dynamics - the Jeans equation. The 
 failure of this method to address the cusp/core issue goes some way to explaining why there is still no consensus. The fundamental limitation of the Jeans equation is a degeneracy between the density 
 profile and the $\beta$ parameter which measures the anisotropy between radial and tangential velocity dispersions. A direct measurement of this quantity is not possible if the observer only has access to the LOS component of each stars velocity and a 2D projection of each stars radius. This is the kinematic data that is available for Milky Way dwarf galaxies at present. As a shorthand we call it the 2+1 scenario as we have access to 2 of the 3 position coordinates and 1 of the 3 velocity coordinates for each star in the sample. 

In practice, the flat LOS velocity dispersion profiles found in real (2+1) dwarf spheroidal data sets 
 are fit equally well \citep{dejonghe92,charbonnier2011} by solutions to the Jeans equation encompassing a huge range of density profiles simply by tuning the anisotropy parameter. More precisely, \cite{Wolf} has recently shown that mass estimates of dwarf 
spheroidal galaxies are approximately independent of the anisotropy parameter $\beta$ at the radius of half-light. If $\beta$ does not undergo sharp transitions within the stellar extent then an inference of the mass at this radius is primarily limited by statistical noise in the velocity measurement. At smaller and larger radii, a free choice of $\beta$ permits a wider range of masses that can fit velocity dispersion measurements. Unless one has some a priori intuition for the anisotropy parameter the mass-anisotropy degeneracy masks the density slope of DM at the galactic center. Under such circumstances, even a broad measurement of the anisotropy parameter (e.g. ruling out $\beta<0.2$) could offer a significant improvement.

\begin{table*}
\caption{Summary of Observational Scenarios: X+Y indicates that we can measure X position coordinates and Y velocity coordinates.}
\centering
\begin{tabular}{c L{12cm}}
\hline
Method & Description  \\
\hline
2+1 & Only the projected radius $R^2=x^2+y^2$ and LOS velocity $v_z$ is available for each star (standard technique). Large data sets (~500+ stars) are currently available for the classical Milky Way dwarf spheroidal population. See the introduction for details of techniques that use 2+1 data.\\
\hline
2+2(3)  & Proper motions of each star are added to $R$ (and $v_z$). As discussed in \cite{evansgaia}, the GAIA satellite's spectrograph has a limiting magnitude (in the G band) of $G\sim 17$ whilst astrometric measurements with microarcsecond precision can be made for stars with $G\sim20$. The GAIA satellite will therefore be able to measure the proper motion of stars before LOS velocities are available in systems such as Galactic globular clusters (at a distance of $\sim$50kpc or less). Mass estimators for this 2+2 scenario are provided in \cite{evansgaia}. See also \cite{wilkoprop} and \cite{strigariprop} for dynamic techniques that can be applied to 2+3 data. In these two works the authors consider the scenario where proper motions can be added to existing LOS velocity measurements in dwarf spheroidal galaxies. \\
\hline
3+1  & In this scenario the LOS depth $z$ is available before the proper motions. This could arise if the variable nature of a star is used to determine its distance from the observer. The LOS depth can be added to $R$ to calculate the deprojected 3D radius $r^2=R^2+z^2$. This is the situation that we investigate in unprecedented detail in this paper. \\
\hline
3+3 & The full 6D information of each star is known. By performing a simple coordinate transformation one has the radial and tangential velocities. The anisotropy parameter $\beta(r)$ can then be read off directly.\\
\hline
\end{tabular}
\label{obssum}
\end{table*}  

There have been several interesting attempts to mitigate the above uncertainty for 2+1 data sets. Here, we provide a brief summary of the simplest analytic methods and guide the interested reader to \cite{batreview} for a 
detailed review of the more sophisticated numerical techniques. With multiple stellar populations, one has multiple half-light radii for which one can estimate the mass.  Given mass estimates at multiple 
points, the mass gradient \citep{penarrubia} can be calculated.  Similarly in the case of Sculptor, \cite{evansvirial} have also exploited multiple stellar populations to place constraints on cusped profiles with 
the Viral theorem.

Without using multiple populations, higher moments of the LOS velocity distribution \citep{Lokas05} can also break the degeneracy between $\beta$ and the DM density profile. A complete generalization of 
the classic Jeans equation analysis has very recently been derived in \cite{me}. The kurtosis (fourth moment divided by the dispersion squared) of LOS velocity data is very sensitive \citep{Lokas02} to the 
anisotropy parameter. By simultaneously fitting to the kurtosis and the velocity dispersions of the data we have an additional constraint on the density profile. This has become an increasingly attractive option 
as dwarf (2+1) data sets have grown. Recent sets of 2+1 kinematic data \citep{walkdat} for Sculptor and Fornax still exceed 1000 stars after applying sophisticated interloper removal schemes. 

In order to obtain maximum information from discrete data however one would like to avoid the binning process altogether and to instead evaluate the likelihood of the fit on a star-by-star basis with the full phase space distribution function $f(\vec{x},\vec{v})$ (see \cite{chakra} for equations that relate the distribution function to observables in numerous X+Y scenarios). In practice, marginalising over all possible distribution functions rather than just fitting to the velocity moments with the Jeans equations presents significant new technical challenges but by utilising Jeans's theorem (i.e. using integrals of motion such as specific energy $E$ and angular momentum $L$ rather than $\vec{x}$ and $\vec{v}$ as coordinates for $f$) the distribution functions are implicitly guaranteed to be in dynamic equilibrium which is particularly convenient when generalising to non-spherical systems. For a more detailed discussion of state-of-the-art discrete modelling techniques we refer the reader to \cite{magorrian}. 

Clearly, if we had all 6D phase space information for stars in a galaxy, we could directly measure the velocity anisotropy $\beta(r)$ with a simple coordinate transform. The degeneracy is broken and an inference of the density profile is only limited by statistical and experimental noise. Short of having all 6D information, as astronomical experiments become ever more sophisticated one might ask what information we need to tackle this degeneracy head on. We summarize the possible observational scenarios in Table~\ref{obssum} and use the notation X+Y to indicate the scenario in which the observer can measure X components of each stars position and Y components of each stars velocity. Estimators for the (global) anisotropy \citep{leonard,anisest2} and the mass \citep{evansgaia} have been developed for 2+3 data sets with projected radii and the full 3D velocity information. Indeed it has been shown \citep{wilkoprop,strigariprop} that the proper motions enable a precise measurement of the density slope (not just the Mass) at the half-light radius. Dwarf galaxies are unfortunately sufficiently 
far away that it will be challenging to get proper motion of stars within a dwarf galaxy in the near future. Proper motions for stars in Galactic globular clusters however may soon be within reach with the GAIA satellite (see Table 1 for details) and this 2+2 scenario is explored in \cite{evansgaia}.    
 
 Alternatively, if we know the depth of stars in the galaxy and LOS motion, we can break the degeneracy between $\beta$ and the slope of the DM profile.
 Depending upon where a star sits in a galaxy, the motion will primarily be either in the tangential or radial direction. Heuristically, from the stars moving in the radial direction we can compute the dispersion in 
 the radial direction and similarly from stars moving in the tangential direction, we can determine the tangential dispersion. 
In the case of the Andromeda galaxy, \cite{watkins} used 500 satellite galaxies  to infer $\beta$ given positions and LOS velocities. We expand upon the technique and apply the more general method 
to dwarf galaxies. 

In the rest of the paper, we focus on the 3+1 method. In Section 2,  we exploit the 3D information to give an intuitive understanding of how $\beta$ is revealed in the LOS velocity data. As a proof of concept,we then 
  explicitly show how we can isolate $\beta$ from the density parameters and thus break the mass-anisotropy degeneracy in a likelihood
  analysis of dwarf spheroidal data. At the end of the section we present very simple estimators for $\beta$ in the form of sample variances. 
  
  In Section 3 the performance of 
  these estimators is tested on mock dwarf spheroidal data from the GAIA challenge. To simulate a population of variable stars in real dwarf spheroidal galaxies we limit our sample size to 500 stars. We show that a sample of 500 stars is not sufficient to precisely determine the local value of 
  $\beta(r)$ but may be used rule out large regions of the parameter space. We also discuss the sensitivity of the estimator to biases from the experimental errors and the 
  assumption of spherical symmetry.   Finally, we turn to the feasibility of our technique.  We find that our technique can likely work for Sagittarius. If observational techniques can be improved by a factor of a few,   our technique may also be applicable to other  nearby dwarfs

\section{Dynamics with 3D positional information}
In this section we show how access to each stars depth along the LOS reveals $\beta$ which parametrizes the anisotropy between radial and tangential velocity dispersions. We show how the LOS depth measurements can be used to tackle the mass-anisotropy degeneracy and derive estimators of $\beta$ that can be applied to dwarf data sets with 3D positions and LOS velocities.

\subsection{The imprint of velocity anisotropy}\label{imprint}

 Let's define the origin of our spherical coordinate system to be at the galaxy's center such that the observer is located a distance $d_c$ in the negative $z$ direction. Assuming that $d_c$ is very large relative to the system's scale (as is the case in dwarf spheroidal galaxies) then we may approximate the distance from the observer to any given star to be $d \simeq d_c+z$.  The projected 2D radius $R^2 = x^2 + y^2$ that is measured on the sky can be combined with z to give the de-projected 3D radius $r^2=x^2+y^2+z^2$ . The line-of-sight velocity of each star is,
\begin{equation}
\label{v_los}
v_z = v_r \cos \theta - v_\theta \sin \theta
\end{equation}  
where $\theta$ is the angle between the star's position vector and the z-axis. Under the assumption of spherical symmetry there is a uniform probability in solid angle $P(\Omega) = 1/4\pi $ and the density of stars $\nu(r)$ depends only on the radius. If the stars belong to an anisotropic distribution function $f(r,\textbf{v})$ then the line-of-sight velocity dispersion at each (deprojected 3D) radius is,
\begin{equation}\label{vz}
\sigma_z^{2}(r) = \frac{1}{\nu}\int \frac{d\Omega}{4\pi} \int d^{3}v  (v_r \cos \theta - v_\theta \sin \theta )^2 f(r,\textbf{v}).
\end{equation}
Performing first the integration over velocity space we use the standard definition,
\begin{equation}\label{moms}
\nu \sigma^{2}_{k}(r) = \int d^{3}v \;v^{2}_{k}f(r,\textbf{v})
\end{equation}
for the velocity dispersions of $f(r,\textbf{v})$ where $k=r,\theta,\phi$. Subsequently integrating over the azimuthal angle $\phi$ and introducing the anisotropy parameter, $\beta = 1 - \sigma^{2}_{\theta}/\sigma^{2}_{r}$ we find,
\begin{equation}\label{vzth}
\sigma^2_z(r,\theta) = \sigma^{2}_{r}(\cos^{2}\theta + (1-\beta) \sin^{2}\theta).
\end{equation}
This result has been derived previously in \cite{leonard}. 

Eq.~\ref{v_los} clearly shows that the motion observed along the LOS depends upon both the radial and tangential velocity. The relative contribution from each component will depend upon the
angle $\theta$ and changes sinusoidally. 

\begin{figure}
        \centering
                \includegraphics[width=9.5cm]{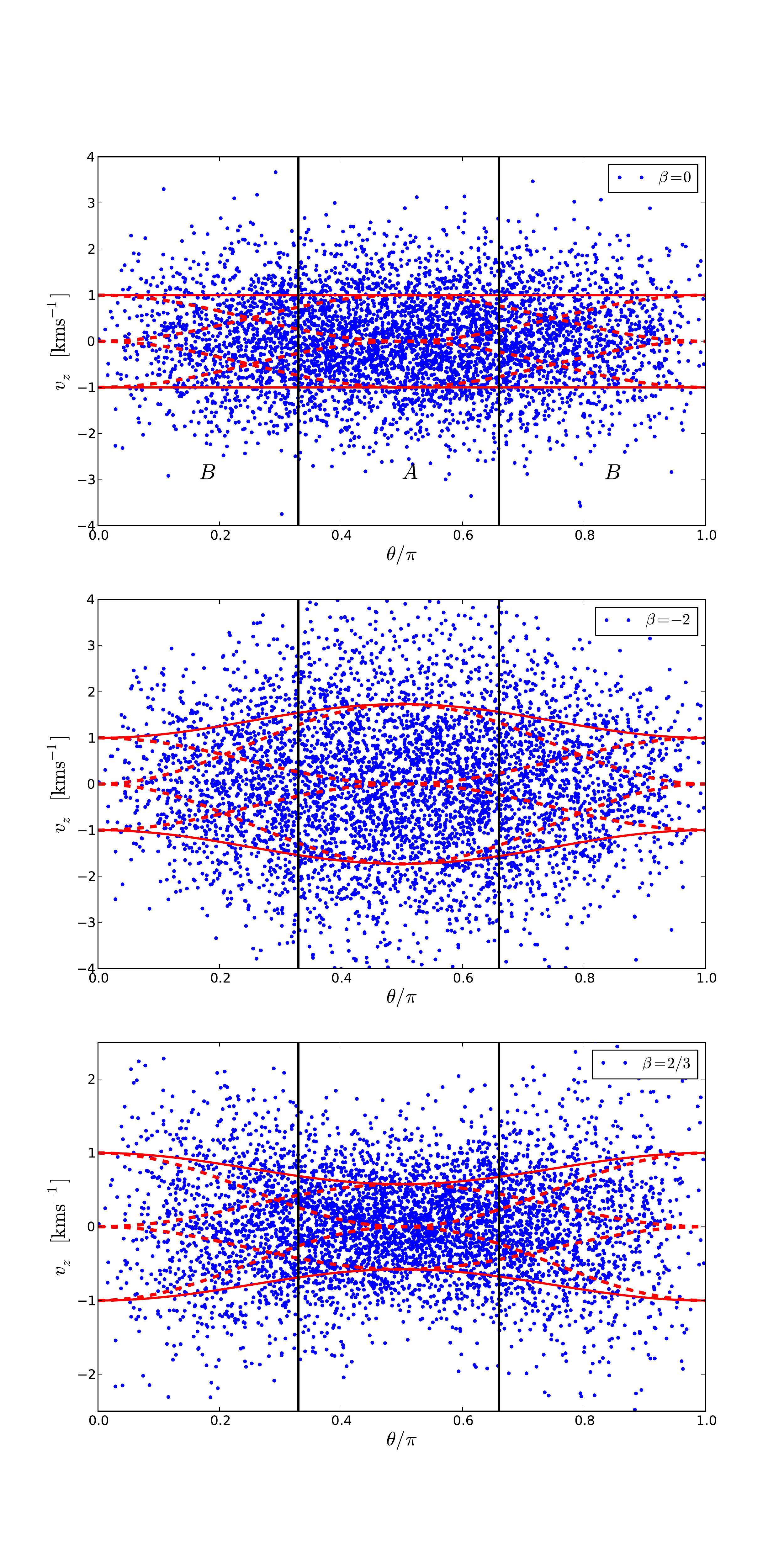}

        \caption{The visual imprint of velocity anisotropy. For systems with isotropic (upper panel), tangentially biased (middle panel) and radially biased (bottom panel) orbits, the LOS velocities of 5000 stars are plotted as a function of the standard polar angle $\theta$ between the position vector of each star and the LOS axis. Dashed red lines show the relative contributions of the radial and tangential dispersions to the LOS dispersion $\sigma_z$ (solid red) as described by Eq. \ref{vzth}. Black verticals lines show the boundaries of regions A and B that are described in Section 2.2.
         }
        \label{visbeta}
\end{figure}

We can visualize the effect. Fig.~\ref{visbeta} plots the LOS velocity of 5000 stars as a function of $\theta$ for a fixed radius $r$.
 The radial velocity dispersion has been set to  $\sigma^2_r = 1$ km/s.
The velocity anisotropy parametrizes a trigonometric envelope which outlines the dispersion of the stars as a function of angle. The shape of this envelope is shown in Fig.~\ref{visbeta} as solid red lines. Dashed red lines show the contribution to the LOS variance from $\sigma^2_r$ and $\sigma^2_\theta$. At the center ($\theta = \pi/2$) of Fig.~\ref{visbeta} the LOS velocities have no radial velocity component and $\sigma_z=\sigma_\theta$. At the edge of the figure ($\theta =0,\pi$) the position vector of each star is aligned with the LOS axis and $\sigma_z=\sigma_r$. 

 With 5000 stars the impact of $\beta$ is visible by eye. In the isotropic case ($\beta=0$), the variance is the same as a function of angle since $\sigma_r=\sigma_\theta$ and the amplitude of the sine and cosine functions are the same. In the tangentially biased case $\beta=-2$, the  variance has a bulge at $\theta = \pi/2$ since $\sigma_r<\sigma_\theta$.
Conversely in the radial biased case $\beta=2/3$, the variance increases at the edges since $\sigma_r>\sigma_\theta$. 

It is also clear from Fig. \ref{visbeta} that if we fix $\sigma_r$ (i.e we fix the solid red lines at the edges of the figure) then the global average of $\sigma^2_z$ over all angles is determined uniquely by $\beta$. The global average of $\sigma^2_z$ increases if $\beta$ is decreased. Averaging Eq. \ref{vzth} over $\theta$ gives     
\begin{equation}\label{sigz}
\sigma^{2}_{z}(r) = \left(1-\frac{2\beta}{3}\right)\sigma^{2}_{r}.
\end{equation}
as was previously derived by  \cite{leonard}.

\subsection{Breaking the mass-anisotropy degeneracy}\label{breakdeg}

In studying dwarf galaxies our primary aim is to constrain the density profile of the dwarf's DM halo. Let's say that we observe a dwarf galaxy and our data set comprises of the 3D positions and LOS velocities of stars. We would like to evaluate the likelihood of the data given a model of the density profile. Typically this is achieved by invoking the assumption of dynamic equilibrium and using the Jeans equation to solve for the variance of the LOS velocity distribution. Unfortunately this is not possible unless we specify the anisotropy parameter $\beta(r)$ in addition to the DM density profile and the density of stars $\nu(r)$. 

Famously there is a degeneracy between the mass and anisotropy parameter if we use the Jeans equation (and subsequent projection for $\sigma^2_z(R)$) to fit LOS dispersion data that is flat as a function of the projected radius $R$. A similar degeneracy persists if we have all three positional components. To break the mass-anisotropy degeneracy with the new LOS depth  measurement we must do more work than simply binning the LOS velocity data in spherical radial bins (as a function of the 3D radius $r$) rather than the usual cylindrical ones. 

Perhaps the simplest and most intuitive way to disentangle the $\beta$ and density parameters is to split the data set into two angular sub-regions (A and B) of equal stellar content. In a spherically symmetric gravitational potential the stars at every angular position have the same contribution from the density parameters. As discussed in Section \ref{imprint} the angular position does however effect the relative contributions of the radial and tangential velocities. We can thus use the angular information to isolate $\beta$ and break the degeneracy. 

From Fig. \ref{visbeta} we see that to maximize the impact of $\beta$ then we should choose region $A$ to be centered at $\pi/2$ and $B$ to enclose the regions nearest the LOS axis in the positive and negative directions. To ensure an equal number of stars in each region (and noting that the stars are distributed uniformly in $\cos\theta$) we define the boundaries at $|\cos\theta|=1/2$. Averaging the LOS dispersion (Eq. \ref{vzth}) over all angles $\theta$ in each sub-region we find,
\begin{equation}\label{sigA}
\langle \sigma^2_z \rangle_A = \left(1-\frac{5\beta}{12}\right)\sigma^2_r\;\;,\;\;\;\;\;\;\;A: |\cos\theta|<\frac{1}{2}
\end{equation}
\begin{equation}\label{sigB}
\langle \sigma^2_z\rangle_B = \left(1-\frac{11\beta}{12}\right)\sigma^2_r\;,\;\;\;\; \;\;B: |\cos\theta|>\frac{1}{2}.
\end{equation}
The ratio of the above equations depends only on $\beta$. We have removed the dependence on the density parameters. By fitting to the velocity dispersion in two angular regions- A and B, we can then solve for $\beta$ and break the degeneracy with the DM profile of the halo.

 In order to test our method, we create a mock dwarf galaxy data set. To model galaxies we often assume that the positions $\{\textbf{x}_i\}$ and velocities $\{\textbf{v}_i\}$ of each tracer star are random samples from a smooth 6D distribution function $f(\textbf{x},\textbf{v})$. We can then parametrize a galaxy by defining the distribution of tracer stars $f$ and the DM density profile $\rho_{\rm{DM}}$. 

 For our purposes here we are only interested in measuring the LOS velocity dispersions.  First, we assume spherical symmetry. Secondly, we also assume for simplicity that the LOS velocity distribution is Gaussian (this assumption will be relaxed in Section 3). With these assumptions, we need only consider the spatial density of stars $\nu(r)$ and the variances of the radial and tangential velocity distributions. As discussed previously these can be determined from the Jeans equation if we specify the DM density profile and the anisotropy parameter $\beta$. The LOS dispersion dispersion then follows from Eq. \ref{sigz}. We therefore define\footnote{To be clear our use of `model' is not a Bayesian model that defines a parameter space but rather one parameter set of a Bayesian model.} a `Model' for a dwarf galaxy by its spatial density of stars $\nu(r)$, its DM density profile $\rho_{\rm{DM}}$ and its anisotropy parameter $\beta$. Every `mock galaxy' discussed in this section is a Monte-Carlo sampling of 5000 stars from a parent Model. 

 We generated a mock galaxy of 5000 stars, galaxy O (for cOre) from parent Model O. To generate each LOS velocity we solved the Jeans equation and used Eq. \ref{vzth} as the variance of a Gaussian LOS velocity distribution. The density of stars $\nu(r)$ in Model O is parametrized by a Plummer profile and has a characteristic radius of 250 pc. The DM distribution of Model O is (as the name suggests) a cored (inner density slope = 0) Hernquist density profile.  The anisotropy parameter of Model O rises from $\beta=0$ at the galactic center to $\beta=1/2$ at large radii. 

Let's say that we are presented with the mock data set galaxy O but don't know that the parent model is the anisotropic and cored Model O. We want to show the power of the 3+1 method to discriminate between Models 
that are degenerate in the standard 2+1 (no LOS depth measurement) Jeans equation analysis. We therefore introduce Model U (for cUsp). Model U has an identical Plummer profile $\nu(r)$ to Model O. The DM profile of Model U is a cusped NFW profile and the anisotropy parameter is $\beta(r)=0$. 

\begin{figure}
        \centering
                \includegraphics[width=9.5cm]{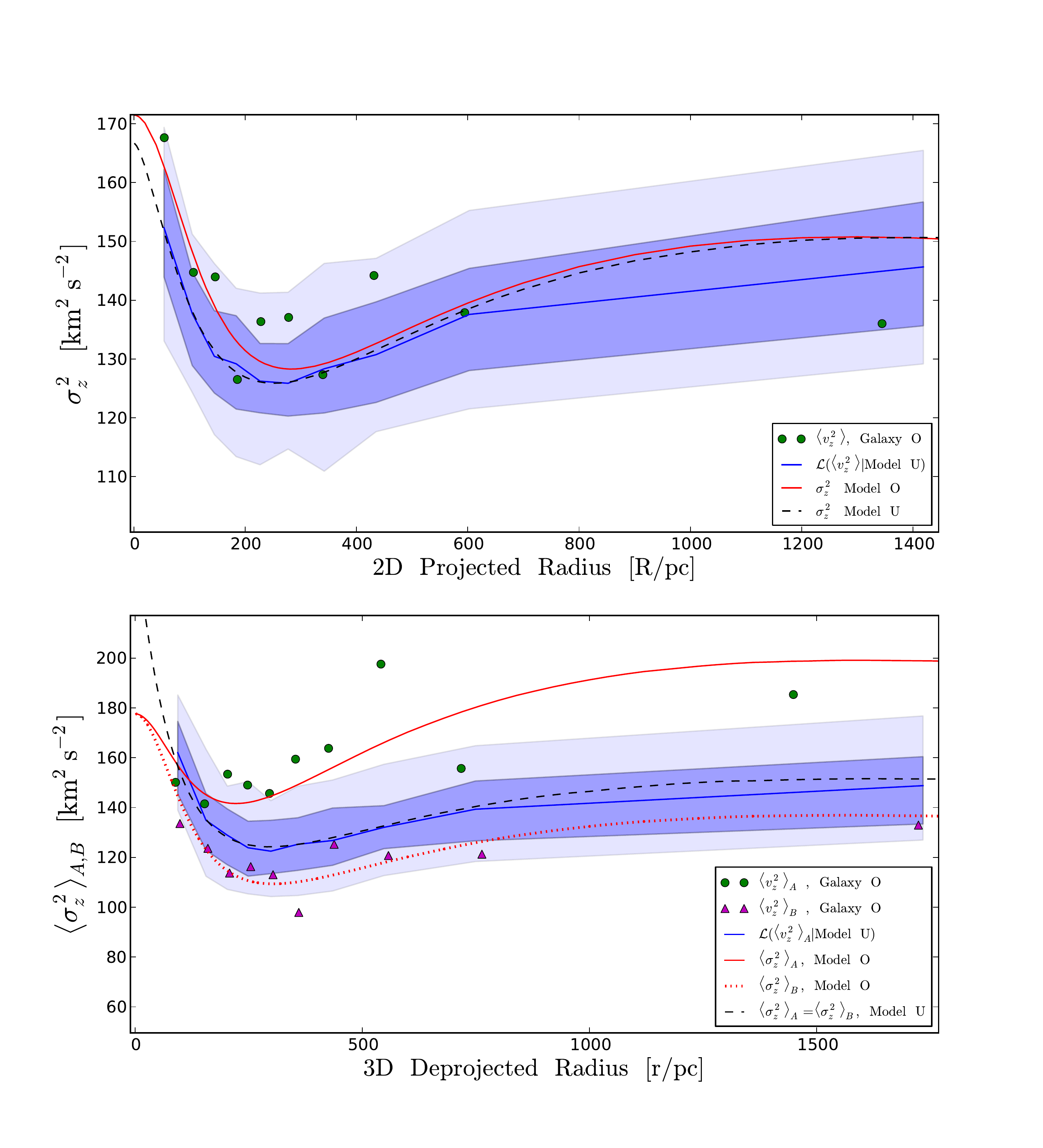}
				\caption{For a mock dwarf galaxy of 5000 stars, Galaxy O, we compare an isotropic ($\beta=0$) model with a cusped density profile (Model U) and an anisotropic model with a cored density profile (Model O). Galaxy O is in fact one Monte-Carlo sampling from Model O (solid and dotted red lines). We show a fit to the data with Model U (dashed black lines). Blue lines and shaded regions show the median and 67 and 95$\%$ contours for the likelihood of the Galaxy O data given the parameters of Model U. See text for details. \protect\\ \textbf{Upper panel} Here we assume that the observer only has access to the 2D projected radius $R$ and the LOS velocities $v_z$. Data points show the sample variance $\langle v^2_z \rangle$ of Galaxy O's LOS velocities in 10 cylindrical radial bins. Solid red and black dashed lines show the Jeans equation prediction for the LOS dispersion after marginalizing over the LOS depth (see eq. 4-57 in \protect\cite{binney87}). \protect\\ \textbf{Lower panel} Now we assume that the observer additionally has the LOS depth measurements of the mock dwarf data. The data may now be divided into two angular sub-regions $A$ and $B$ according to the condition $|\cos\theta|<1/2$. Each half of the Galaxy O data set is then split into 10 spherical radial bins. Green circular and magenta triangular data points show the sample variance of LOS velocities in regions $A$ and $B$. Red solid and dotted lines show the expected values of the LOS dispersion in each angular region according to Eqs. \ref{sigA} and \ref{sigB}. We note that that as Model U is isotropic then $\langle \sigma^2 \rangle_A=\langle \sigma^2\rangle_B$. For clarity we therefore only show the likelihood contour for the $\langle v^2_z \rangle_A$ data.}       
        \label{massdeg}
\end{figure}

 In the top panel of Fig.~\ref{massdeg}, we assess the standard (2+1) method on galaxy O and demonstrate the mass-anisotropy degeneracy. In this case we only have access the projected radii $R$ and LOS velocities of the 5000 stars in galaxy O. The data points in the upper panel show the LOS velocity dispersion $\langle v^2_z \rangle$ of galaxy O's stars in 10 cylindrical radial bins. The solid red and dashed black lines show the variance of the LOS velocity distribution $\sigma^2_z$ in Model O and U respectively. Despite the very different DM density profiles of Models O and U we see by eye that by suitable tuning of the anisotropy parameter $\beta$ the LOS variances are very similar.

 To assess the statistical significance of this claim for galaxy O we must compute the likelihood. We can perform a frequentist estimate for the likelihood of a Model by generating many mock galaxies from the model parameters and determining the regions where the median and central 67 and 95$\%$ of our estimators (in this case $\langle v^2_z \rangle$) lay. The blue region shows the resulting likelihood contours for Model U. Clearly with only 5000 stars the width of the contours is significantly larger than the difference between the two models. Though we do not explicitly show the likelihood given the cored Model O, it is evident that the likelihood of the data given our cusped isotropic model is very similar to that of our cored model with 
tuned anisotropy. We are thus unable to reject either hypothesis. 

With the 3+1 method we can break the degeneracy.
The lower panel of Fig.~\ref{massdeg} shows what we see if we additionally measure the LOS depth of each star. By splitting galaxy O into angular sub-regions $A$ and $B$, we see that the LOS velocity dispersions in each 
region ( region A- green circular and region B- magenta triangular) are visibly divided at large radii. This splitting of dispersions in region A and B is a smoking gun for anisotropy $\beta \neq 0$ as can be seen in Fig.~\ref{visbeta}. In particular we see from the bottom panel of Fig. \ref{visbeta} that if the LOS dispersion in region A is larger than in region B then this indicates radial anisotropy $\beta>0$. The solid and dotted red lines show the variance of the LOS velocity distribution in each region for the anisotropic Model O. Conversely the Model U will have the same LOS velocity distribution in both angular regions which is shown with a black dashed line. 

Hence, Model U can not simultaneously fit the LOS dispersion data of galaxy O in angular region A and B. The degeneracy between the two models is broken. The blue likelihood contours (in this case the estimator is $\langle v^2_z \rangle_A$) for Model U show that this is a statistically significant result for our mock galaxy of 5000 stars.   

Of course 5000 stars is an unrealistically large sample size that is chosen for illustrative purposes. We note in Fig. \ref{massdeg} that we may also have ruled out tangentially anisotropic models (where $\langle \sigma^2_z \rangle_A$ is instead smaller than $\langle \sigma^2_z \rangle_B$) with an even greater significance. Even for modest sample sizes where a precise measurement of the splitting is obscured by noise it should still be possible to place a strong upper or lower bound.                 

\subsection{Estimators for $\beta$}

In the previous section we described how one could directly tackle the degeneracy problem. For a real galaxy it may often be the case that only a small subset of available stars are amenable to a precise measurement of the LOS depth. Additionally this measurement will be subject to a myriad of potential uncertainties that introduce complicated systematic biases. In either case it may therefore be preferable to use the limited 3D data set to make a more robust estimate $\beta$ and to use this as a prior in existing 2+1 methods that treat larger and more precisely measured 2D data sets. In this section we introduce simple estimators for $\beta$.

Firstly let's adapt the method applied in Section \ref{breakdeg}. Eliminating the $\sigma^2_r$ term we derive the estimator for $\beta$,
\begin{equation}\label{betest2}
\widehat{\beta}_0 = 12\frac{\langle v^2_z \rangle_B -\langle v^2_z \rangle_A}{11\langle v^2_z\rangle_B-5\langle v^2_z \rangle_A}
\end{equation}
that approximates $\beta$ for a radial bin of stars. Though useful as an illustrative example this method requires a splitting of the data that increases the statistical noise by a factor of approximately $\sqrt{2}$. We can instead obtain a new constraint by reweighting the line-of-sight velocities by $\cos^{2}\theta$ and again performing the average over solid angles, 
\begin{eqnarray}\label{vzcs}
\overline{\sigma^{2}_{z}\cos^{2}\theta} &=& \int \frac{d\Omega}{4\pi} (\sigma^{2}_{r} \cos^{4}\theta + \sigma^{2}_{\theta}\sin^{2}\theta \cos^{2}\theta)\\
&=& \frac{1}{3}\left(1-\frac{2\beta}{5}\right)\sigma^{2}_{r}.
\end{eqnarray}
Again we may cancel out $\sigma^2_r$ to get a new estimator,
\begin{equation}\label{betest}
\widehat{\beta}_1 = \frac{3}{2}-\frac{1}{5\omega-1}, \;\;\; \omega \equiv \frac{\langle v^{2}_{z}\cos^{2}\theta\rangle}{\langle v^{2}_{z}\rangle}.
\end{equation}
that uses the full data set. 

As an aside one could also perform a maximum likelihood analysis on the individual data points to derive $\beta$ but we found that introducing the angular dependence strongly altered the simple Gaussian velocity distributions. Numerical tests indicated that the maximum likelihood method is more susceptible to the complicated biases that are discussed in the next section.   
\section{Monte-Carlo performance of $\beta$ estimators for Mock Dwarf Data}

In this section, we use mock dwarf spheroidal data from the GAIA challenge workshop (accessible on the wiki- \cite{gaiawiki}) to test the variance and bias of our anisotropy parameter estimators 
(Eqs. \ref{betest2} and \ref{betest}). 
We draw stars from a more realistic distribution function $f(r,\textbf{v})$ rather than the Gaussian approximation used in the preceding sections. 
 As a point of comparison, the fake data sets have been drawn from two different but realistic models. Both models have identical Plummer stellar density profiles with a characteristic radius of $250$ pc embedded in a DM halo with a scale radius of 1 kpc. 

 Model 1 has a cusped NFW density 
 profile and an isotropic velocity distribution at all radii ($\beta=0$). Model 2 has a cored DM halo and Osipkov Merritt anisotropy where $\beta$ rises from zero at the galactic 
 center to purely radial orbits ($\beta=1$) at large radii. For more details on these models we refer the reader to the GAIA Challenge data suite described in \cite{gaiamodel}. These two models were selected as they have very different anisotropy parameters but similar flat dispersion profiles.

Dwarf galaxies may have on the order of 10 thousand plus stars. In the feasibility section we will focus on variable stars as our tracer population.  Dwarf galaxies typically only have several hundred variable stars so we limit our sample size to $N=500$.   

A rigorous treatment of uncertainties in the distance measurement is a complex and method-dependent issue that is beyond the scope of this work. Indeed studies (See  \cite{rave} for a treatment of stars in the RAVE survey \citep{ravedata}) of how to quantify the error distribution of distances to stars are ongoing. As such we consider the performance of our estimator for a simplified model in which the errors for the  distance measurements are assumed to be Gaussian before commenting on the impact of a more realistic error distribution. 

To evaluate the variance and bias of our anisotropy estimators we drew a large number of independent samples of $N=500$ stars from the GAIA Challenge data set. To these positions and velocities we added Gaussian distributed noise to the LOS depth $z$ and velocity $v_z$ measurement with standard deviations $\delta_z$ and $\delta_{vz}$ and centered at the true coordinates. For each sample we split the data into 5 radial bins of 100 stars, calculated the anisotropy estimator for each bin and added it to an array. In this way we were able to construct likelihood regions for $\widehat{\beta}$ in a similar fashion to those for $\langle v^2_z \rangle$ in Fig. \ref{massdeg}. 

\begin{figure}
        \centering
                \includegraphics[width=8.5cm]{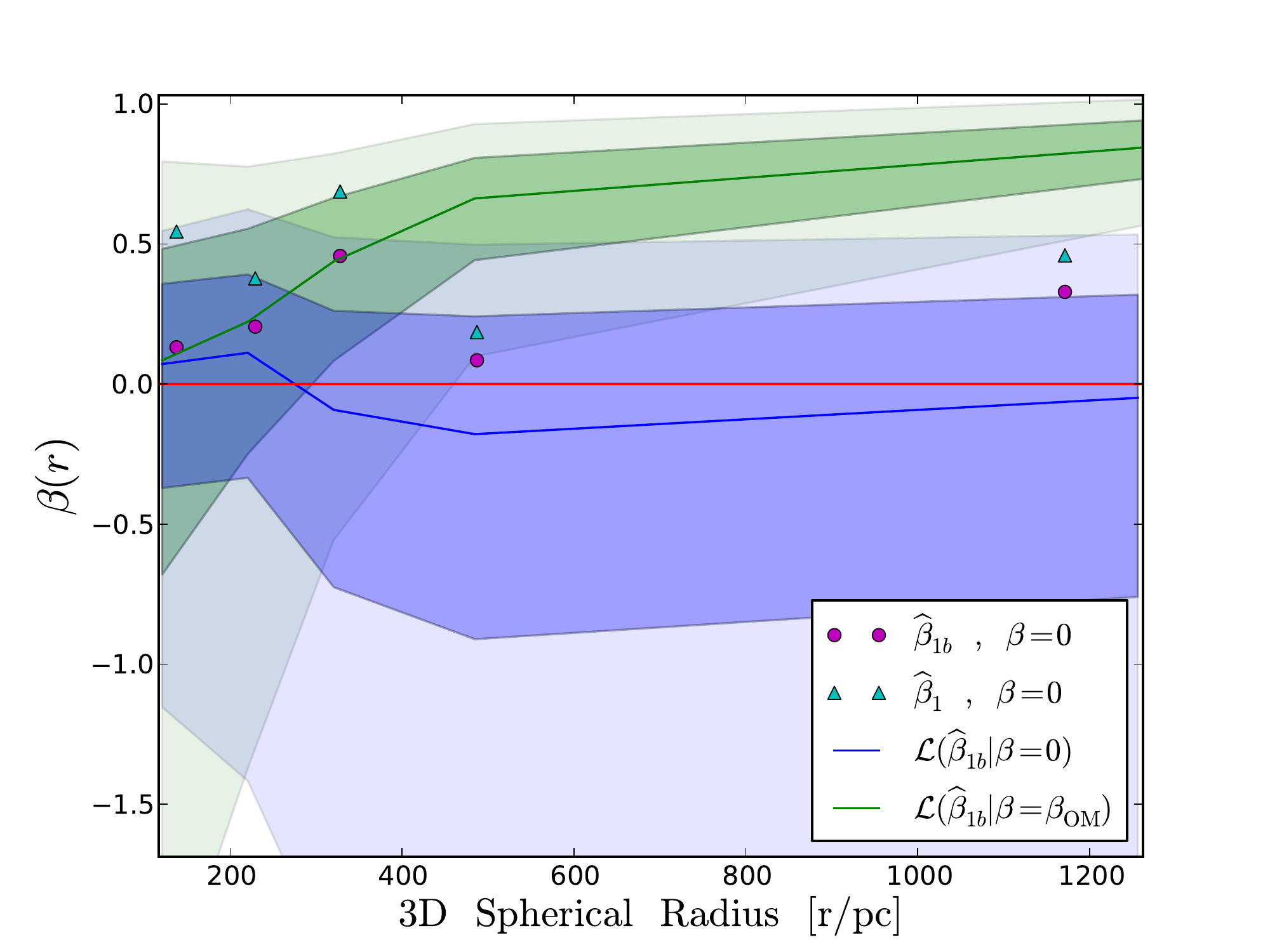}
     \caption{Five hundred stars were sampled from a mock dwarf galaxy with an isotropic distribution function ($\beta=0$ solid red line). Gaussian errors were added to the LOS depth and velocity to simulate experimental errors. Data points show the anisotropy parameter estimator $\widehat{\beta}_{1b}$ (bias corrected) for five spherical radial bins of 100 stars.  
       Shaded regions show the likelihood of the data points given the true isotropic model (blue) and a model with Osipkov-Merritt anisotropy (green) where $\beta$ undergoes a transition from zero at the galactic center to 
      one at large radii about transition radius $r_a=250$pc. Central lines show the median and shaded regions show the central $67$ and $95\%$ intervals.}
        \label{betaplot}
\end{figure}

\subsection{Sources of bias and an improved estimator $\widehat{\beta}_{1b}$}
For a robust inference of $\beta$ and in particular to claim a stringent upper or lower bound we must consider how experimental errors and other systematics bias our estimators.
 The experimental errors will introduce biases in our estimate for $\beta$. 

Our numerical tests indicate that only substantially large LOS velocity errors $\delta_{vz} \geq 8$ km/s can bias our estimators $\widehat{\beta}$. Adding Gaussian distributed noise to the LOS velocities adds a systematic error to the LOS dispersion that is independent of the position of the star. In Fig. \ref{visbeta} we see that this would simply shift the red lines vertically. Because the estimators $\widehat{\beta}$ are defined as ratios of the LOS dispersion measurements at different positions, this systematic effectively cancels out. In practice velocity measurements should be sufficiently precise that we can neglect the bias due to velocity.     

Instead, we focus upon the experimental errors in the LOS depth measurement,
which will alter the distribution of stars first morphologically and secondly in terms of ordering.  These two effects will introduce two separate biases.

Morphologically, experimental distance errors may artificially elongate the galaxy along the LOS. In our Gaussian example with width $\delta_z$, the square average of the LOS depth measurement $z$ is transformed via $\langle z^2 \rangle \to \langle z^2 \rangle + \delta^2_z$. Measurements of $x$ and $y$ in plane perpendicular to the LOS are left unchanged however so we have $\langle z^2 \rangle > \langle x^2 \rangle=\langle y^2\rangle$. In other words our spherically symmetric halo becomes ellipsoidal. 
As our estimators are derived under the assumption of spherical symmetry (and particularly uniformity in solid angle) the elongation will bias the inferred value of $\beta$. For the interested reader, the bias is similar to the Lutz \citep{lutz} bias. In the Lutz case however, Gaussian errors in the parallax measurement ($p \propto 1/d$) have the effect of squashing the galaxy. 
  
The depth measurement errors will introduce a second bias when we group the stars into radial bins. For sufficiently large scatter, the ordering of stars will be corrupted by moving stars from one bin to another, which introduces a separate ordering bias effecting $\widehat{\beta}$. Though the two aforementioned biases can have a dramatic effect on $\widehat{\beta}$ we now demonstrate that there are hints from the data that allow us to identify them and in some cases clean them.

 First, we introduce a new estimator to remove the morphological bias for our estimate of $\beta$. We can identify the departure from spherical symmetry in our data set by measuring  
\begin{equation}
b = \frac{1}{N}\frac{\sum^{N}_{i} \cos^{2}\theta_i}{\langle \cos^2\theta \rangle_{\rm{true}}} = \frac{3}{N} \sum^{N}_{i} \cos^{2}\theta_{i}.
\end{equation}
A spherical system with uniform probability in solid angle has $\langle \cos^{2}\theta \rangle=1/3$. The parameter $b$ indicates the fractional deviation from this prediction. By design, the parameter $b$ enables us to formulate a simple estimator for the anisotropy parameter, 
\begin{equation}
\widehat{\beta}_{1b} = \frac{3}{2}-\frac{b}{5\omega-b}
\end{equation} 
for which we can correct for the morphological bias. 
In Fig. \ref{betaplot}, we have plotted our original estimator $\widehat{\beta}_1$ without the correction. We adopt Gaussian errors of $\delta_z=100$pc and ($\delta_{vz} = 3$kms$^{-1}$) which are shown in 
Fig. \ref{betaplot}. These errors  are significant relative to the the half-light radius $R_{1/2} = 250$pc and the 
LOS dispersion $\sigma_z \approx 11$km/s.  Clearly, the original estimator is offset from the true value of $\beta=0$. Conversely, our morphologically corrected estimator $\widehat{\beta}_{1b}$ now aligns 
with the true value of $\beta$. The solid blue line in Fig. \ref{betaplot} that represents the median value of $\widehat{\beta}_{1b}$ for a large sample of isotropic mock data sets is approximately centered at 
zero. We found that if we increased the distance errors to much above $\delta_z = 150$pc then even our improved estimator $\widehat{\beta}_{1b}$ began to show significant bias. 

Our corrected estimator assumes that all of the deviation from spherical symmetry is a result of the experimental errors. If the system is not spherically symmetric we would need to introduce a different 
estimator with an alternative prediction for $\langle \cos^2\theta\rangle$. This could be guided by evaluating the deviations from spherical symmetry that are present in the $x$ and $y$ positional data in the 
plane perpendicular to the LOS. In other words we can check to see that the galaxy looks circular in projection as is the case in Draco. Any such analysis would vary from dwarf to dwarf and we leave this more  complicated case for the future. 

 We can also predict the effects of the ordering bias, which mixes the stars from different radial bins. Though we measure the position of a star at a point with observed radius $r_o$, the true position of the star is actually at some other point with a true radius $r_t$.  Given our limited knowledge, we will never know $r_t$. If we know the density of stars at each radius $\nu(r)$ and the distribution of errors in the depth measurement then we can use a likelihood analysis to determine the probable 
 location, which will allow us mitigate the ordering bias. The 3D density of stars can be estimated by fitting to the surface density of stars that we observe on the sky and performing an Abel inversion.

Let's denote the error in the radial measurement $\alpha_r$ such that $r_o=r_t+\alpha_r$. If we know the distribution of LOS depth errors $\alpha_z$ then we may use
\begin{equation}
\alpha_r = \sqrt{r^2_t+r_t\cos\theta\alpha_z+\alpha_z^2}-r_t
\end{equation}
to determine the distribution of radial measurement errors $P(\alpha_r|r_t)$.  In our example where the LOS depth errors $\alpha_z$ are normally distributed we can easily derive the displacement error distribution function $P(\alpha_r|r_t)$ numerically by 
drawing $\alpha_z$ from the normal distribution and $\cos\theta$ uniformly from -1 to 1. Even though we have a fixed error distribution for $\alpha_z$ we note that $P(\alpha_r|r_t)$ depends on the true radius $r_t$. We are now in a position to evaluate the expectation value of the true radius $\langle r_t \rangle$ at an observed radius $r_o$. This is simply a weighted combination of all $r_t$ and $\alpha_r$ that combine to make the observed radius $r_o$, namely, 
\begin{eqnarray}
\langle r_t \rangle (r_o) &=& \int^{\infty}_{0}dr N(r) r \int^{\infty}_{-\infty} d\alpha_r P(\alpha_r|r) \delta(r_o-r-\alpha_r)  \nonumber \\ 
&=& \int^{\infty}_{0}dr N(r) r P(\alpha_r=r_0-r|r)
\end{eqnarray}
where the delta function ensures that $r_0 = r + \alpha_r$. Our $\beta$ estimators depend on sample variances of the LOS velocities. With errors in the LOS depth we must update our estimates of the LOS dispersion to reflect the fact that the majority of stars at $r_o$ originated from radii centered at $\langle r_t \rangle $. The impact of this bias depends on gradients in the LOS dispersion. If the gradients are large, then $\sigma^2_z(r_o)$ could be considerably different to $\sigma^2_z(\langle r_t \rangle)$. Fortunately real dwarf spheroidals have approximately flat LOS dispersion profiles which minimizes this effect.  

In summary, the estimators are prone to bias for any realistic account of the distance errors.  Though this bias is not trivial to calculate we have shown with our simple model that if we have an understanding of the error distributions and a good estimate of the stellar density profile then there is sufficient information from the data to identify the bias and in some cases to clean it. We found that the estimators are insensitive to errors in the LOS velocity measurement but that even our improved estimator $\widehat{\beta}_{1b}$ can not account for bias if the error in the LOS depth measurement climbs above $\delta_z=150$pc.

  In reality the biggest issue that the method faces is that the estimator is sensitive to the assumption of spherical symmetry which is already in tension with observations of real dwarf spheroidals. Sadly the only motivation for this assumption is simplicity. Nonetheless the spherical model provides a useful test example to answer a more general question. If we can only observe the LOS component of each velocity then how well can we utilize the full 3D positional information to determine the velocity anisotropy? For any configuration of stellar positions, the key to answering this question lies in determining the average contribution of each velocity coordinate to the LOS.       

\subsection{How well could we determine $\beta$ in practice?}

The primary concern of this section is the fundamental limitation of sample size. In practice distance measurements of stars in dwarf spheroidal galaxies will be limited to variable stars. This will be discussed in greater detail in the feasibility section. We therefore consider small samples of 500 stars which represents the upper limit of variable star populations in dwarf spheroidals. We again adopt the benchmark errors of $\delta_{vz}=3$kms$^{-1}$ and $\delta_z = 100$pc in the LOS velocity and depth. In the previous section we showed that these errors are not sufficient to bias the estimator $\widehat{\beta}_{1b}$.
 
 From Fig. \ref{betaplot} we note that with 500 stars the blue likelihood contours corresponding to the isotropic galaxy are broad at all radii offering little distinction between data points in areas of mild radial ($\widehat{\beta}>0$) and tangential ($\widehat{\beta}
<0$) anisotropy. As with the estimator $\widehat{\beta}_0$ this reflects the fact that isotropy does not leave a strong signal on the angular distribution. By contrast we see from the green 
contours that if the galaxy is strongly anisotropic, as is the case for the Osipkov-Merritt model at large radii, then the observed data points will almost always be clustered 
tightly about the true value. This is a key result. With a small sample of 500 tracers the strength of the 3+1 method depends on the nature of the galaxy that we observe. 

 If the galaxy that we observe is approximately isotropic then clearly the statistical noise limits an inference of $\beta$. The scatter of data points will be consistent with a wide range of mildly anisotropic models and an inference is limited to setting upper and lower bounds that rule out more extreme models. Such information would still be of interest.  Mass estimates at the half-light radius \citep{Wolf} are less robust if the anisotropy parameter has large gradients. The mass slope method \citep{penarrubia} in particular could be sensitive to such a correction if individual sub-populations have contrasting gradients. Though intuition suggests that large gradients in $\beta$ are unlikely it would be useful to validate this assumption with observation. Any prior information on $\beta$ could also be useful for methods that use higher moments (i.e that measure the shape as well as the width) of the velocity distribution. As discussed in the introduction, these methods are much more sensitive to $\beta$ than the classic Jeans analysis of velocity dispersions. 
 
 If the galaxy that we observe is instead highly anisotropic then we can make even stronger assertions. Had we sampled our mock galaxy from the anisotropic Osipkov-Merritt model then we see from the green contours in Fig. \ref{betaplot} that 95$\%$ of measurements of $\beta_{1b}$ at large radii ($r>1000$pc) will be constrained above $\widehat{\beta}_{1b}=0.5$. The blue contours show that an observation of $\widehat{\beta}_{1b}=0.5$ is in tension with the isotropic model at the $2\sigma$ level. Of course, with one data point this is not highly significant but it can be used to place much more stringent constraints on tangential models with $\beta<0$. If the model was highly anisotropic at all radii then 5 observations of $\widehat{\beta}_{1b}=0.5$ would be sufficient to exclude the isotropic model. 

In this favorable scenario then limiting the parameter space to positive $\beta$ with the 3+1 method can break the degeneracy if combined with existing (2 projected positions + 1 LOS velocity) methods. The dwarf that we observe will have a much larger (we no longer require that the stars are variable) complementary data set of 2D projected radii plus LOS velocities. The left panel of fig. $1$ in \cite{Wolf} shows the degenerate solutions of the Jeans equation that fit the (typically) flat LOS velocity dispersion of the Carina dSph. The mass at the half-light radius is fixed but the mass at the center is masked by the mass-anisotropy degeneracy. In particular we see that (in the case of constant anisotropy) radially anisotropic models with smaller interior masses (i.e cores) are degenerate with tangentially anisotropic models with large interior masses (i.e cusps). The prior of $\beta>0$ set by our 3+1 method can thus break the degeneracy between cusps and cores.  
 
\section{Feasibility}

There are several techniques to determine the distance of individual stars in dwarf galaxies ranging from parallax to variable stars.  
We discuss how the different distance measurement techniques could be useful. See Table~\ref{obssum} for definitions of the different methods.
We find that variable stars offer the best opportunity to apply the 3+1 method. First we consider parallax, which turns out to be more important for methods other 
the 3+1 method.  With sufficiently good angular position 
and long enough baseline, we could get all 6D phase space information for a galaxy. 
Hence, one would use all the data and not be limited to a 3+1 method.  
 Furthermore even without sufficiently good parallax data to determine the depth of a star along the LOS,
 we could still get the proper motion of the stars inside the galaxy. 
The proper motion of stars in a Dwarf plus the LOS velocity would allow one to use the 2+3 method discussed in the introduction. Hence, parallax would be more useful for other techniques.

In the near term, proper motion studies will be limited.  
To measure the internal motion of stars inside a dwarf will require an observe to determine the proper motion to the percent level. 
Gaia in the next 5 years will map out the proper motion and position for over a billion stars and will significantly alter our understanding of the Milky Way.  
Regardless, Gaia  will have difficulty to achieve the necessary precision (percent level) to observe the internal motion of stars inside of Dwarf galaxies, with the possible exception of Sagittarius.

Ground based missions and pointing telescopes maybe able to do better than Gaia.
In the early 1950's, \cite{1961AJ.....66..300B} made a detail study of the  variable stars in Draco.
Over an entire season Baade took many long and deep  exposures  (100 plates plus)
to determine the period of nearly 300 RR Lyrae.  Subsequently over the last 20 years , many telescopes including (CFHT, HST, KPNO, USNO, etc) have imaged Draco and UMiI
with CCDs.   Following a technique discussed and used by \cite{2012AAS...21925217S,2013A&A...554A.101B}, 
one should be able to use at least a 20 year baseline to determine the proper motion and position of the stars in Draco to the few percent level. A longer base line (60 years)
could reduce the error to the sub-percent level.   Hence Draco and Sagittarius are potentially interesting targets for the 2+3 method and may even allow for a full 6D analysis. Regardless,
parallax data will be sparse and hard to come by in the near term.

Given the potential challenges of parallax, we look to other methods to determine the depth of a star along the LOS.
Variable stars are very promising.  Unlike proper motion studies, one does not need a long baseline to infer a distance. In addition,
dwarf galaxies typically have several hundred variable stars such as RR Lyrae.
Fornax has at least 500 RR Lyrae \citep{2002AJ....123..840B}, 
Draco has nearly  300 hundred RR Lyrae  \citep{2008AJ....136.1921K} if not more, and the Sagittarius dwarf has at least 400 stars \citep{2011AcA....61....1S}.
 In the case of RR Lyrae stars, one can certainly  determine
the absolute distance of RR Lyrae stars to the percent level using a Period Luminosity Relationship (PLR) \citep{2013ApJ...776..135M}.  
 Much of the error in the absolute distance is due to the constant off set term of the PLR.
We are only interested in the relative depth of each star in the dwarf galaxy.  Error in the off set term will not alter the relative depth of a star
with respect to the center of the galaxy.  Hence, the 3+1 method will not be affected by errors related to the absolute distance relationship.

The fundamental limitation of the PLR for a dwarf spheroidals would be due to any intrinsic scatter.
In the V band, this is certainly true \citep{1990ApJ...350..603S}. 
In the K band, theoretical models of the PLR \citep{2004ApJS..154..633C,2001MNRAS.326.1183B} appear to have an intrinsic scatter of about 0.032 mags,
which corresponds to about a 2\%$\sim\delta d/d$  uncertainty in relative distance (where d is the distance to the star and $\delta d$ is the error in the distance).
\cite{2001MNRAS.326.1183B} assumed an intrinsic mass scatter of about 4 \% which accounts for most of the intrinsic scatter of the PLR by contributing 0.03 mags.  
If one can account for the mass, 
one could reduce any intrinsic scatter. 

The K band PLR depends upon metallicity. For simplicity, we have neglected any uncertainties in metallicity in the intrinsic scatter.
First, we  can directly measure the metallicity of RR Lyrae stars. Second, dwarfs such as Draco do not appear to not have a large range of metallicities. 
On the positive side, Kepler has brought new life into the field with new exquisite light curves for RR Lyrae \citep{2013ASSP...31..109K}.  
Kepler's new insight has improved modeling of RR Lyrae, which may also have the side benefit of improving the PLR.

Regardless, we would like to observe directly any intrinsic scatter in the PLR.  
Unfortunately at present, observational errors mask any intrinsic scatter.  
\cite{1992MmSAI..63..433B} looked at stars in M3 and $\omega$ Cen. The authors found a scatter of about 0.03 mags, 
which appears to be consistent with limited number of observations.  
If the scatter found by \cite{2004ApJS..154..633C,2001MNRAS.326.1183B} is real, then 
the mass and metallicity of stars in  M3 and $\omega$ Cen  are almost identical.  Conversely, the PLR relationship may actually be  less sensitive 
to the mass and metallicity of stars than has been found by   \cite{2004ApJS..154..633C,2001MNRAS.326.1183B}.
In sum at present, the PLR relationship appears limited to the percent error level in determining distances, but the relationship might still be improved 
in the future.

Beyond the PLR,  the Baade-Wesselink (BW) method~\citep{1926AN....228..359B,1946BAN....10...91W, 1969MNRAS.144..297W}
 may also be used to directly measure the distance to RR Lyrae in dwarf galaxies.
The BW method   measures the change 
in a star's luminosity as a function of the star's surface velocity and temperature, from which one can directly infer the intrinsic luminosity.
The BW method is limited by how well we can measure the temperature of the star and the star's actual surface velocity.

Even without recent improvements in stellar modeling and various means to determine the temperature of a star such as with widths of metal lines,
the temperature of RR Lyrae stars have previously been determined down to 20 K with only K and V band photometry (see \citep{1992ApJ...396..219C} for further references).
\cite{2010arXiv1007.3441B,2012MNRAS.419.2774B} has formulated a new way to directly model the size mass, radius and temperature of the star by
exploiting multiple bands- No spectral information. The authors could determine the temperature to within 20 K and limit the mass to better than 4 percent. We emphasis that the conclusions drawn by
\cite{2010arXiv1007.3441B,2012MNRAS.419.2774B} were made without the benefit of
deep infrared band coverage (H, J, and K) and without the use of any spectral information. 
In terms of surface velocity,
we note that 
with the addition of spectral information, we can measure the expansion of a star down to a fraction of a km/s.   Apogee can measure
the velocity of stars down to 50 m/s~\cite{apogee}.  The stars in the Dwarfs will be much fainter than the stars observed with apogee (mag 20 vs. mag 13), but we can
use a 10 meter telescope  versus a 1 meter telescope, which will give a much faster integration time. Regardless, we probably can still measure the surface velocity with 
a precision on the order of few tenths of a km/s.
With more probes (multiple bands and spectroscopy), we may well be able to improve our determination of the surface temperature, mass, and radius of RR Lyrae.   
More complimentary data can not hurt.

As noted before, {our numerical tests on mock dwarf spheroidal data indicate that} the 3+1 method requires that we know the relative distance to better than
150 pc. We can translate that into a relative distance error of $\delta d/d$ where $d$ is the distance to the
star. We take $\delta d$=150 pc. For a fixed $\delta d$, we can then infer the maximum error allowed to apply the 3+1 method.
For example, the maximum error allowed for Sagittarius is 0.7\%,   0.3\% for Ursa Minor, 0.2\% for Draco, and 0.1\% for Fornax. In summary we will need distances to a few tenths of a percent.

A factor of a few improvement in the PLR and the BW method would be sufficient to apply the 3+1 method to the above mentioned galaxies.
As noted previously, PLR  appears to only determine the relative distance down to the percent level (multiple bands and measurement
of the stars metallicity could improve the situation). As noted previously, any inherent dispersion appears is either limited to taking insufficient data 
or due to mass dispersion. If one can use multiple bands to better constrain the mass of RR Lyrae then, PLR maybe sufficient to apply the 3+1 method.

  How well can the BW method work?
The temperature of an RR Lyrae star
is ($T\sim$7000 K).  The star has a surface velocity of ($V\sim$ 150 km/s).  As a base line, we will assume an error $\delta V$ of .2 km/s and  $\delta T$ of 20 K,
which gives a relative error on the distance $\delta d/d$ to around 0.7\% with $\delta d/d\simeq 2\delta T /T+\delta V/V$ (we have neglected
uncertainties on the period-which is small). Most of the error comes from the temperature. Remarkably even without an improvement of BW method
we could already successfully apply the 3+1 method to Sagittarius. 
 A factor of a few improvement in the determination of temperature would allow us to also apply the 3+1 method to  Ursa Minor, Draco and possibly Fornax. 
Finally, we emphasis that even with the present limitations on the BW method, the 3+1 method could be used to help constrain $\beta$ in Sagittarius.

\section{Discussion}
Without knowledge of the anisotropy parameter $\beta$, the standard methods in galactic dynamics are not able to unambiguously 
 determine the DM density profile at the center of the halo. By adjusting $\beta$, one can accommodate a cored or a cusped profile.
 For systems as distant as dwarf spheroidal galaxies this degeneracy can't be broken at present because the available LOS velocities and projected radii offer virtually no 
information on the anisotropy. With new sophisticated astronomical techniques this could be subject to change in the near future.

The scenario in which data sets are bolstered by the proper motions has been examined thoroughly in the literature. For observational reasons however less attention has been paid to the 
case where we first gain access to the LOS depth of each star. In this work we illustrate clearly how velocity anisotropy can be identified with this information and show for spherically 
symmetric systems how one could in theory break the mass-anisotropy degeneracy completely. For an application to small samples we developed simple estimators for the anisotropy 
parameter $\widehat{\beta}$ that could be used to inform priors on $\beta$ in existing 2D methods. This would be particularly effective for methods that employ higher moments of the velocity 
distribution which are sensitive to the anisotropy parameter.  

 To gauge how our estimators might perform in practice we tested them on realistic mock dwarf data from the GAIA Challenge. Samples of $N=500$ stars were given Gaussian experimental 
 errors of $\delta_z = 100$ pc in the LOS depth and $\delta_{vz} = 3$ km/s in the LOS velocity measurement. With so few stars we found that a precise measurement of $\widehat{\beta}$ is 
 only possible for highly anisotropic galaxies that leave a strong imprint on the data. The performance of the method is thus dependent on the nature of the galaxy. 

 If the galaxy of interest is approximately isotropic ($\beta=0$) then the statistical noise limits an inference of $\beta$ to setting upper and lower bounds. Given that we effectively have no 
 a priori intuition for $\beta$ this is still useful information and excluding models with radical anisotropy would verify the implicit assumptions of widely used mass estimators. 

The results are much more dramatic if the galaxy is in fact highly anisotropic. In this case the estimators $\widehat{\beta}$ are clustered tightly about the true value. Our numerical tests with 
500 stars show that an observation of $\widehat{\beta}=0.5$ is in tension with tangential models $\beta<0$ at the $2\sigma$ level. If such an observation were combined with the large 2D data 
sets available for dwarf spheroidals then the degeneracy can be broken.

 In numerical tests with simple Gaussian error distributions we found that the estimators $\widehat{\beta}$ are robust to errors in the LOS velocity measurements but prone to bias from 
 uncertainties in the distance measurements. For our test model dwarfs with a half-light radius of 250pc the bias became highly significant as distance errors climbed above 150pc. This is only 
 a guide and it is beyond the scope of this work to fully quantify the bias as in practice the relevant error distributions will dependent on the observational method. Where possible we have 
 briefly shown how one can use the data to identify and clean elements of this bias. 

An additional issue is the assumption of spherical symmetry which is in tension with measurements of real Milky Way dwarf spheroidals. Numerical tests must be conducted to quantify how 
well the estimators fare on non-spherical halos but the spherical model nevertheless gives us an interesting insight into the opportunities that the 3D positional information provides. The 
general idea holds for non-spherical stellar distributions; with 3D positional data we can see exactly how much each velocity component contributes to a stars line-of-sight. This varies locally 
and by splitting or reweighting the data according to its position we can isolate the velocity anisotropy and break the degeneracy.  

Finally,  we argue that our method could successfully be applied to Sagittarius. Our numerical tests on mock dwarf spheroidal data indicate that we must know the relative depth of stars in a galaxy to within 150 pc. As noted in the feasibility section, dwarf galaxies typically have a population of several hundred RR Lyrae. Via the Baade-Wesselink (BW) method we can measure the distance to the star directly. The BW method is sensitive to accurate measurements of the surface temperature and the radial velocity of the stars surface. With a highly accurate spectrograph and coverage with multiple bands (especially with the addition of the K band), it may well be possible to apply our method directly to Sagittarius. If we can improve the inferred temperature of an RR Lyrae  by a factor of 3 or 4, we could also apply the method to Draco, Ursa Minor, and potentially other nearby dwarf galaxies.

\section*{Acknowledgments}
TR thanks the KCL Graduate School and Malcolm Fairbairn for support and Paolo Gondolo and Celine Boehm for useful discussions. 
He would also like to thank Jorge Penarrubia and Matthew Walker for making their mock dSph data publicly available. 
The research of D.S. has been supported at IAP by  the ERC project  267117 (DARK) hosted by Universit\'e Pierre et Marie Curie - Paris 6.
We would like to acknowledge the Mainz Institute for Theoretical Physics (MITP) for enabling us to complete a significant portion of this work.

\bibliographystyle{mn2e2}
\bibliography{TR7}

\end{document}